\documentclass{ws-procs9x6}

\begin{document}

\title{COSMOGRAPHIC ANALYSIS OF DARK ENERGY}

\author{MATT VISSER and C\'ELINE CATTO\"EN}

\address{School of Mathematics, Statistics, and Operations Research\\
Victoria University of Wellington\\
New Zealand\\
E-mail: \{matt.visser,celine.cattoen\}@msor.vuw.ac.nz\\
http://msor.victoria.ac.nz/}

\begin{abstract}
The Hubble relation between distance and redshift is a purely cosmographic relation that depends only on the symmetries of a FLRW spacetime, but does not intrinsically make any dynamical assumptions. This suggests that it should be possible to estimate the parameters defining the Hubble relation without making any dynamical assumptions. To test this idea, we perform a number of inter-related cosmographic fits to the {\sf legacy05} and {\sf gold06} supernova datasets, paying careful attention to the systematic uncertainties.  Based on this supernova data, the ``preponderance of evidence'' certainly suggests an accelerating universe. However we would argue that (unless one uses additional \emph{dynamical} and \emph{observational} information, and makes additional \emph{theoretical assumptions}) this conclusion is not  currently supported ``beyond reasonable doubt''.  As part of the  analysis we develop two particularly transparent graphical representations of the redshift-distance relation --- representations in which acceleration versus deceleration reduces to the question of whether the relevant graph slopes up or down. 
\end{abstract}

\keywords{Cosmography, Hubble parameter, deceleration parameter, jerk.}

\bodymatter

\section{Introduction}
\def\fl{} 
\def\eg{\emph{e.g.}}

When analyzing the case for ``dark energy'', it is critically important to realize that 
the standard luminosity distance versus redshift relation~\cite{Weinberg, Peebles},
\begin{eqnarray}
d_L(z) =  {c\; z\over H_0}
\Bigg\{ 1 + {\left[1-q_0\right]\over2} {z} 
+ O(z^2) \Bigg\},
\label{E:Hubble1a}
\end{eqnarray}
and its higher-order extension~\cite{Chiba, Sahni, Jerk, Jerk2},
\begin{equation}
d_L(z) =  {c \; z\over H_0}
\Bigg\{ 1 + {\left[1-q_0\right]\over2} {z} 
-{1\over6}\left[1-q_0-3q_0^2+j_0+ {kc^2\over H_0^2\,a_0^2}\right] z^2
+ O(z^3) \Bigg\},
\label{E:Hubble1}
\end{equation}
are purely cosmographic results applicable to \emph{any} FLRW universe, regardless of the assumed dynamics. Following the spirit of Hubble's original proposal~\cite{Hubble1929}, one could in principle  fit such a relation directly to the supernova data~\cite{legacy, legacy-url, gold, Riess2006a, Riess2006b, essence}, thereby estimating cosmological parameters (such as $H_0$, $q_0$, and the jerk $j_0$) without making any dynamical assumptions ---  but we shall see that there are ways of pre-processing the Hubble relation to make the result (and potential problems) stand out in greater clarity~\cite{Hubble-arXiv, Hubble-PRD, Hubble-CQG, Hubble-CQG2, Hubble-JCAP}.

For instance, it is sometimes observationally more convenient to count photons rather than energy, and consider the ``photon flux distance''~\cite{Hubble-arXiv, Hubble-PRD, Hubble-CQG}
\begin{equation}
 d_F = {d_L\over(1+z)^{1/2}},
\end{equation} 
for which, defining
\begin{equation}
d_H = {c\over H_0}; 
\qquad\hbox{and}\qquad
\Omega_0=1+{kc^2\over H_0^2a_0^2} = 1 + {k \; d_H^2\over a_0^2};
\end{equation}
one derives
\begin{equation}
d_F(z) =  {d_H z }
\Bigg\{ 1 - {q_0 {z} \over2}
+{\left[3+10q_0+12q_0^2-4(j_0+\Omega_0)\right]\over24} z^2
+ O(z^3) \Bigg\}.
\end{equation}
Furthermore, using the ``distance modulus'' in terms of which the supernova data is actually reported~\cite{legacy, legacy-url, gold, Riess2006a, Riess2006b, essence}
\begin{equation}
\mu_D = {5} \; \log_{10}[d_L/(10 \hbox{ pc})] = {5} \; \log_{10}[d_L/(1 \hbox{ Mpc})] +25,
\end{equation}
one has the simple relation
\begin{equation}
\ln[d_F/(z \hbox{ Mpc})] = {\ln10\over5} [\mu_D - 25] - \ln z - {1\over2} \ln(1+z),
\end{equation}
leading to a particularly useful form of the Hubble relation:
\begin{equation}
\ln\left[{d_F\over z \hbox{ Mpc}}\right] 
= \ln\left[{d_H\over\hbox{Mpc}}\right]
 - {q_0  {z} \over2}
+{\left[3+10q_0+9q_0^2-4(j_0+\Omega_0)\right]\over24} z^2
+ O(z^3).
\end{equation}
Note that the question of whether or not the universe is accelerating or decelerating now reduces to the simple question of whether or not the curve obtained by plotting $\ln[d_F/(z \hbox{ Mpc})] $ versus $z$ slopes up or down.

\begin{figure}[!htbp]
\begin{center}
\includegraphics[width=9.5cm]{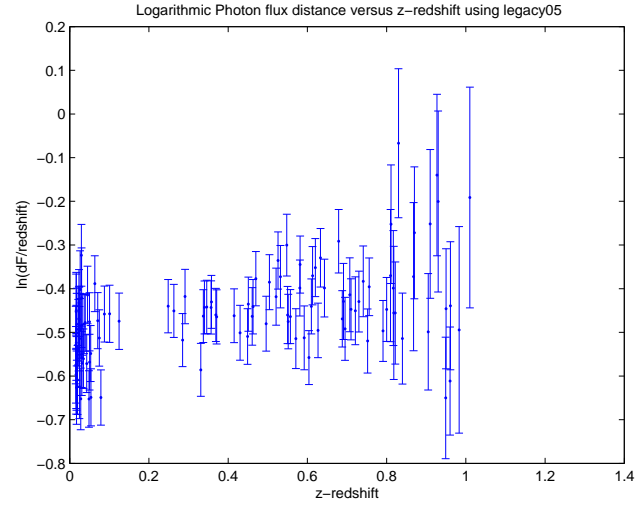}
\end{center}
\caption{\label{F:ln_dF_legacy05}
The normalized logarithm of the photon flux distance, $\ln(d_F/[z \hbox{ Mpc}])$, as a function of the $z$-redshift using the {\sf legacy05} dataset~\cite{legacy, legacy-url}. As is traditional in the field, the plotted error bars do not include estimates of the systematic errors.}
\end{figure}

\begin{figure}[!htbp]
\begin{center}
\includegraphics[width=9.5cm]{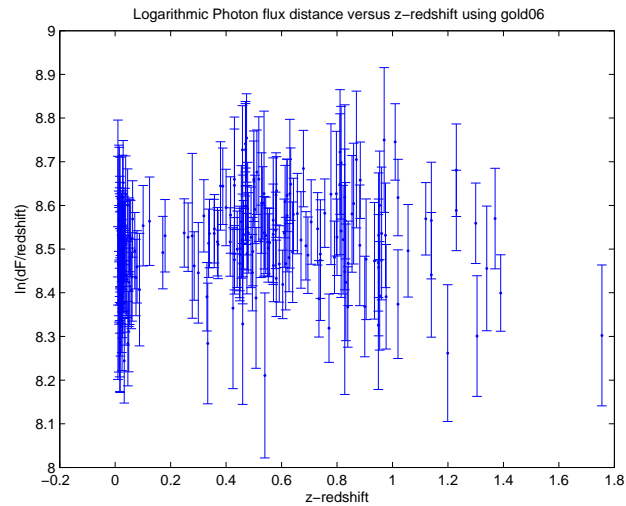}
\end{center}
\caption{\label{F:ln_dF_gold06}
The normalized logarithm of the photon flux distance, $\ln(d_F/[z \hbox{ Mpc}])$, as a function of the $z$-redshift using the {\sf gold06} dataset~\cite{gold, Riess2006a}. As is traditional in the field, the plotted error bars do not include estimates of the systematic errors.}
\end{figure}

For the data plots presented in this article we have used data from the supernova legacy survey ({\sf legacy05})~\cite{legacy,legacy-url} and the Riess \emph{et.~al.}~``gold'' dataset of 2006 ({\sf gold06})~\cite{Riess2006a}.  (The {\sf gold06} dataset is a larger dataset that contains most but not all of the {\sf legacy05} supernovae.) Note that figures \ref{F:ln_dF_legacy05} and \ref{F:ln_dF_gold06} are not plots of ``statistical residuals'' obtained after curve fitting --- rather they can be interpreted as plots of ``theoretical residuals'', obtained by first splitting off the linear part of the Hubble law (which is now encoded in the intercept with the vertical axis), and secondly choosing the quantity to be plotted so as to make the slope of the curve at redshift zero particularly easy to interpret in terms of the deceleration parameter.  

The plots presented in figures \ref{F:ln_dF_legacy05} and \ref{F:ln_dF_gold06} are considerably more ambiguous than we had initially expected. In generating these plots and performing the statistical analysis to be described below (considerably more detail can be found at~\cite{Hubble-arXiv, Hubble-CQG, Hubble-PRD}) we had initially hoped to verify the robustness of the Hubble relation, and to possibly obtain improved estimates of cosmological parameters such as the deceleration and jerk parameters, thereby complementing other recent cosmographic and cosmokinetic analyses such as~\cite{Blandford,  Blandford0, Shapiro, Caldwell, Elagory}, as well as other analyses that take a sometimes skeptical view of the totality of the observational data~\cite{paddy1, paddy2, paddy3, paddy4, barger}.   

In view of the rather disturbing visual impact of  figures \ref{F:ln_dF_legacy05} and \ref{F:ln_dF_gold06} we resolved to see if they could be improved by further transformations of the data. For instance, we looked at the possibility of transforming the redshift variable, we looked at the possibility of adopting a number of other distance surrogates, and we performed a detailed statistical analysis of the data paying careful attention to the question of estimating the systematic uncertainties~\cite{Hubble-arXiv}. While the ``preponderance of evidence'' certainly suggests an accelerating universe, we would argue that (unless one uses additional \emph{dynamical} and \emph{observational} information, and makes additional \emph{theoretical} assumptions) this conclusion is not currently supported ``beyond reasonable doubt''.  The supernova data (considered in isolation) certainly \emph{suggests} an accelerating universe, but it is not sufficient to allow us to reliably conclude that the universe \emph{is} accelerating.\,\footnote{From recent discussions with a broad cross section of the community, it appears that this result now seems to have become part of the standard ``folklore''. Typically, statistically strong arguments for cosmic acceleration rely on working within a particular dynamical framework (such as $\Lambda$CDM), and on extra observational data (such as independent constraints on $\Omega_0$ coming from CMB observations). }

\section{New redshift variable: $y=z/(1+z)$}

Because much of the recent supernova data is being acquired at large redshift ($z\gtrsim1$), there are a number of theoretical reasons why it might be more appropriate to adopt the modified redshift variable~\cite{Hubble-arXiv, Hubble-PRD, Hubble-CQG}
\begin{equation}
y = {z\over1+z}; \qquad z={y\over1-y}.
\end{equation} 
In the past (of an expanding universe)
\begin{equation}
z \in (0,\infty); \qquad  y\in (0,1);
\end{equation}
while in the future
\begin{equation}
z \in (-1,0); \qquad  y\in (-\infty,0).
\end{equation}
Thus the variable $y$ is both easy to compute, and when extrapolating back to the Big Bang has a nice finite range $(0,1)$. Furthermore, Taylor series in terms of the $y$ variable have improved convergence properties at high redshift~\cite{Hubble-arXiv, Hubble-PRD, Hubble-CQG}. We will refer to this variable as the \emph{$y$-redshift}.~\footnote{Similar expansion variables have certainly been considered before. See, for example,  Chevalier and   Polarski~\cite{Polarski}, who effectively worked with the dimensionless quantity $b=a(t)/a_0$, so that $y=1-b$. Similar ideas have also appeared in several related works~\cite{Linder, Linder2, Bassett, Martin}. Note that these authors have typically been interested in parameterizing the so-called $w$-parameter, rather than specifically addressing the Hubble relation.}
In terms of the variable $y$:
\begin{equation} \label{dL}
\fl
d_L(y) =  {d_H\; y}
\Bigg\{ 1 - {\left[-3+q_0\right]\over2} {y} 
+{\left[12-5q_0+3q_0^2-(j_0+ \Omega_0) \right]\over6} y^2
+ O(y^3) \Bigg\}.
\end{equation}
It is now useful to define a quantity
\begin{equation}
 d_Q = {d_L\over(1+z)^{3/2}} = {d_F\over 1+z} = (1-y) \; d_F,
\end{equation} 
which we shall refer to as the ``deceleration distance''. This quantity has the nice feature that
\begin{equation}
\ln[d_Q/(y \hbox{ Mpc})] = {\ln10\over5} [\mu_D - 25] - \ln y + {3\over2} \ln(1-y),
\end{equation}
whence
\begin{equation}
\ln\left[{d_Q\over y \hbox{ Mpc}}\right] = \ln\left[{d_H\over \hbox{Mpc}}\right]
 - {q_0\,  {y} \over2} 
+{\left[3-2q_0+9q_0^2-4(j_0+\Omega_0)\right]\over24} y^2
+ O(y^3).
\end{equation}
Thus plotting $\ln[d_Q/(y \hbox{ Mpc})]$ versus $y$ results in a curve whose slope at redshift $y=0$ is directly proportional to the deceleration parameter:  The question of whether or not the universe is accelerating or decelerating now reduces to the simple question of whether or not the curve obtained by plotting $\ln[d_Q/(y \hbox{ Mpc})] $ versus $y$ slopes up or down.

\begin{figure}[!htbp]
\begin{center}
\includegraphics[width=9.5cm]{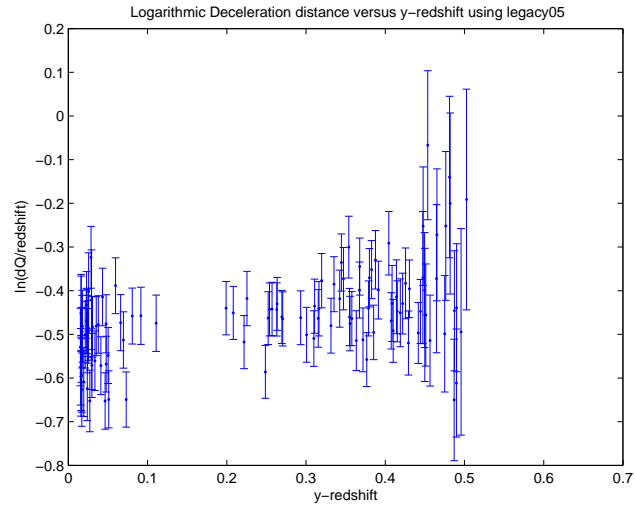}
\end{center}
\caption{\label{F:ln_dQ_legacy05}
The normalized logarithm of the deceleration distance, $\ln(d_Q/[y \hbox{ Mpc}])$, as a function of the $y$-redshift using the {\sf legacy05} dataset~\cite{legacy, legacy-url}. As is traditional in the field, the plotted error bars do not include estimates of the systematic errors.}
\end{figure}

\begin{figure}[!htbp]
\begin{center}
\includegraphics[width=9.5cm]{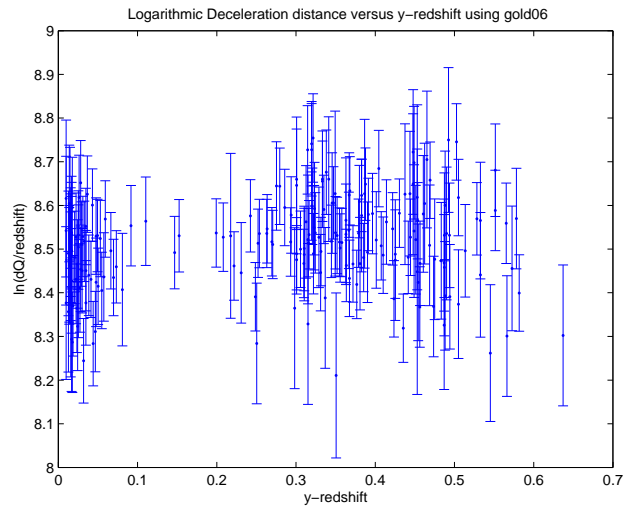}
\end{center}
\caption{\label{F:ln_dQ_gold06}
The normalized logarithm of the deceleration distance, $\ln(d_Q/[y \hbox{ Mpc}])$, as a function of the $y$-redshift using the {\sf gold06} dataset~\cite{gold, Riess2006a}. 
As is traditional in the field, the plotted error bars do not include estimates of the systematic errors.}
\end{figure}

Visually, the plots presented in figures \ref{F:ln_dQ_legacy05} and \ref{F:ln_dQ_gold06} are again considerably more ambiguous than we had initially expected. Note that up to this point we have not performed any statistical analyses, we have ``merely''  found a dramatic way of visually presenting the observational data.

\section{Data fitting: Statistical tests}

In view of the somewhat ambiguous and possibly  alarming nature of the plots presented in figures  \ref{F:ln_dF_legacy05}--\ref{F:ln_dQ_gold06},   we then performed a number of statistical tests and analyses to check the extent to which robust and reliable results could be obtained from the supernova data considered in isolation. (For extensive technical details see~\cite{Hubble-arXiv, Hubble-PRD}, and related theoretical discussion in~\cite{Hubble-CQG}). In performing the statistical analyses reported below we compared and contrasted results using several notions of cosmological distance, and two different versions of redshift. The distance surrogates we used were~\cite{Hubble-arXiv,Hubble-CQG, Hubble-PRD}:
\begin{itemize}
\item The ``luminosity distance'' $d_L$.

\item The ``photon flux distance'': $d_F = {d_L\;(1+z)^{-1/2}}$ .

\item The ``photon count distance'':  $d_P = {d_L\;(1+z)^{-1}}$.

\item The  ``deceleration distance'': $d_Q = {d_L\;(1+z)^{-3/2}}$.

\item  The ``angular diameter distance'': $d_A = {d_L\;(1+z)^{-2}}$.

\end{itemize}
The $z$-based versions of the Hubble law we used were~\cite{Hubble-arXiv,Hubble-CQG}:
\begin{eqnarray}
&&\ln\left[{d_L\over z \hbox{ Mpc}}\right] = {\ln10\over5} [\mu_D - 25] - \ln z
\\
&& \quad = \ln\left[{d_H\over \hbox{Mpc}}\right]
\nonumber
 - {\left[-1+q_0\right]\over2} {z} 
+{\left[-3+10q_0+9q_0^2-4(j_0+\Omega_0)\right]\over24} z^2
+ O(z^3).
\nonumber
\\[4pt]
&&\ln\left[{d_F\over z \hbox{ Mpc}}\right] = {\ln10\over5} [\mu_D - 25] - \ln z - {1\over2} \ln(1+z)
\\
&& \quad = \ln\left[{d_H\over \hbox{Mpc}}\right]
 - {q_0  {z}\over2}
+{\left[3+10q_0+9q_0^2-4(j_0+\Omega_0)\right]\over24} z^2
+ O(z^3).
\nonumber
\\[4pt]
&&\ln\left[{d_P\over z \hbox{ Mpc}}\right] = {\ln10\over5} [\mu_D - 25] - \ln z -  \ln(1+z)
\\
&& \quad = \ln\left[{d_H\over \hbox{Mpc}}\right]
 - {\left[1+q_0\right]\over2}  {z} 
+{\left[9+10q_0+9q_0^2-4(j_0+\Omega_0)\right]\over24} z^2
+ O(z^3).
\nonumber
\\[4pt]
&& \ln\left[{d_Q\over z \hbox{ Mpc}}\right] = {\ln10\over5} [\mu_D - 25] - \ln z - {3\over2} \ln(1+z)
\\
&& \quad = \ln\left[{d_H\over \hbox{Mpc}}\right]
 - {\left[2+q_0\right]\over2}  {z} 
+{\left[15+10q_0+9q_0^2-4(j_0+\Omega_0)\right]1\over24} z^2
+ O(z^3).
\nonumber
\\[4pt]
&&\ln\left[{d_A\over z \hbox{ Mpc}}\right] = {\ln10\over5} [\mu_D - 25] - \ln z - 2 \ln(1+z)
\\
&& \quad = \ln\left[{d_H\over \hbox{Mpc}}\right]
 - {\left[ 3 + q_0\right]\over2}  {z} 
+{\left[21+10q_0+9q_0^2-4(j_0+\Omega_0)\right]\over24} z^2
+ O(z^3).
\nonumber
\end{eqnarray}
%
Similarly, the $y$-based versions of the Hubble law we used were~\cite{Hubble-arXiv,Hubble-CQG}:
\begin{eqnarray}
&&\ln\left[{d_L\over z \hbox{ Mpc}}\right] = {\ln10\over5} [\mu_D - 25] - \ln y
\\
&& \quad = \ln\left[{d_H\over \hbox{Mpc}}\right]
 -{\left[-3+q_0\right]\over2} {y} 
+{\left[21-2q_0+9q_0^2-4(j_0+\Omega_0)\right]\over24} y^2
+ O(y^3).
\nonumber
\\[4pt]
&&\ln\left[{d_F\over z \hbox{ Mpc}}\right] = {\ln10\over5} [\mu_D - 25] - \ln y + {1\over2} \ln(1-y)
\\
&& \quad = \ln\left[{d_H\over \hbox{Mpc}}\right]
 - {\left[-2+q_0\right]\over2}  {y} 
+{\left[15-2q_0+9q_0^2-4(j_0+\Omega_0)\right]\over24} y^2
+ O(y^3).
\nonumber
\\[4pt]
&&
\ln\left[{d_P\over z \hbox{ Mpc}}\right] = {\ln10\over5} [\mu_D - 25] - \ln y +  \ln(1-y)
\\
&& \quad = \ln\left[{d_H\over \hbox{Mpc}}\right]
 - {\left[-1+q_0\right]\over2}  {y} 
+{\left[9-2q_0+9q_0^2-4(j_0+\Omega_0)\right]\over24} y^2
+ O(y^3).
\nonumber
\\[4pt]
&&\ln\left[{d_Q\over z \hbox{ Mpc}}\right] = {\ln10\over5} [\mu_D - 25] - \ln y + {3\over2} \ln(1-y)
\\
&& \quad = \ln\left[{d_H\over \hbox{Mpc}}\right]
 - {q_0\over2}\,  {y} 
+{\left[3-2q_0+9q_0^2-4(j_0+\Omega_0)\right]\over24} y^2
+ O(y^3).
\nonumber
\\[4pt]
&&\ln\left[{d_A\over z \hbox{ Mpc}}\right]= {\ln10\over5} [\mu_D - 25] - \ln y + 2 \ln(1-y)
\\
&& \qquad = \ln\left[{d_H\over \hbox{Mpc}}\right]
 - {\left[ 1 + q_0\right]\over2}  {y} 
+{\left[-3-2q_0+9q_0^2-4(j_0+\Omega_0)\right]\over24} y^2
+ O(y^3).
\nonumber
\end{eqnarray}
Fits were carried out for all five distance surrogates, and for both definitions of redshift, using polynomial approximants to the Hubble relation up to 7th-order.~\footnote{Note that because the uncertainty in the redshift is encoded in the uncertainty of the distance modulus, the uncertainty in logarithmic distance is just scaled by a factor of ln(10)/5. Therefore, if the uncertainty is gaussian in the distance modulus, it is also gaussian in logarithmic distance, which is crucial for least squares fitting.} The $F$-test was then used to discard statistically meaningless terms, and it was seen that quadratic fits were the best that could meaningfully be adopted.~\footnote{Note that in a cosmographic framework, where one is most closely following the spirit of Hubble's original methodology~\cite{Hubble1929}, one does not have a dynamical model to fit the data to, and the use of least-squares fits to a truncated Taylor series is the best one can possibly hope for.  Ultimately, the truncated Taylor series method is not really a very radical approach, being firmly based in quite standard statistical techniques~\cite{basic1, basic2, basic3, basic4, transformation, nonlinear}.} The results are presented in tables \ref{T:q_y_legacy05}--\ref{T:q_z_gold06}. 
\enlargethispage{10pt}
\begin{table}[!htdp]
\caption{ } 
\begin{center}
{Deceleration and jerk parameters ({\sf legacy05} dataset, $y$-redshift).}\\[5pt]
\begin{tabular}{|c|c|c|}
\hline
distance & $q_0$ & $j_0+\Omega_0$ \\
\hline
$d_L$ & $-0.47\pm 0.38$ & $-0.48\pm3.53$ \\
$d_F$ &  $-0.57\pm0.38$ & $+1.04\pm3.71$ \\
$d_P$&  $-0.66\pm 0.38$& $+2.61\pm3.88$\\
$d_Q$ &  $-0.76\pm0.38$& $+4.22\pm4.04$\\
$d_A$ &  $-0.85\pm0.38 $& $+5.88\pm4.20 $\\
\hline
\end{tabular}
\\[5pt]
{\small With 1-$\sigma$ statistical uncertainties.}
\end{center}
\label{T:q_y_legacy05}
\vskip-10pt
\end{table}%
\begin{table}[!htdp]
\caption{ } 
\begin{center}
{Deceleration and jerk parameters ({\sf legacy05} dataset, $z$-redshift).}\\[5pt]
\begin{tabular}{|c|c|c|}
\hline
distance & $q_0$ & $j_0+\Omega_0$ \\
\hline
$d_L$ & $-0.48\pm 0.17$ & $+0.43\pm0.60$ \\
$d_F$ &  $-0.56\pm0.17$ & $+1.16\pm0.65$ \\
$d_P$&  $-0.62\pm 0.17$& $+1.92\pm0.69$\\
$d_Q$ &  $-0.69\pm0.17$& $+2.69\pm0.74$\\
$d_A$ &  $-0.75\pm0.17 $& $+3.49\pm0.79 $\\
\hline
\end{tabular}
\\[5pt]
{\small With 1-$\sigma$ statistical uncertainties.}
\end{center}
\label{T:q_z_legacy05}
\vskip-10pt
\end{table}%
\begin{table}[!htdp]
\caption{ } 
\begin{center}
{Deceleration and jerk parameters ({\sf gold06} dataset, $y$-redshift).}\\[5pt]
\begin{tabular}{|c|c|c|}
\hline
distance & $q_0$ & $j_0+\Omega_0$ \\
\hline
$d_L$ & $-0.62\pm 0.29$ & $+1.66\pm2.60$ \\
$d_F$ &  $-0.78\pm0.29$ & $+3.95\pm2.80$ \\
$d_P$&  $-0.94\pm 0.29$& $+6.35\pm3.00$\\
$d_Q$ &  $-1.09\pm0.29$& $+8.87\pm3.20$\\
$d_A$ &  $-1.25\pm0.29 $& $+11.5\pm3.41 $\\
\hline
\end{tabular}
\\[5pt]
{\small With 1-$\sigma$ statistical uncertainties.}
\end{center}
\label{T:q_y_gold06}
\vskip-10pt
\end{table}%
%
\begin{table}[!htdp]
\caption{ } 
\begin{center}
{Deceleration and jerk parameters ({\sf gold06} dataset, $z$-redshift).}\\[5pt]
\begin{tabular}{|c|c|c|}
\hline
distance & $q_0$ & $j_0+\Omega_0$ \\
\hline
$d_L$ & $-0.37\pm 0.11$ & $+0.26\pm0.20$ \\
$d_F$ &  $-0.48\pm0.11$ & $+1.10\pm0.24$ \\
$d_P$&  $-0.58\pm 0.11$& $+1.98\pm0.29$\\
$d_Q$ &  $-0.68\pm0.11$& $+2.92\pm0.37$\\
$d_A$ &  $-0.79\pm0.11 $& $+3.90\pm0.39 $\\
\hline
\end{tabular}
\\[5pt]
{\small With 1-$\sigma$ statistical uncertainties.}
\end{center}
\label{T:q_z_gold06}
\end{table}%

Even after we have extracted these numerical results there is still a considerable amount of interpretation that has to go into  understanding their physical implications~\cite{Hubble-arXiv, Hubble-PRD}. In particular note that the differences between the various models, (Which distance do we use? Which version of redshift do we use? Which dataset do we use?), often dwarf the statistical uncertainties within any particular model.  If better quality (smaller scatter) data were to become available, then one could hope that the cubic term would survive the $F$-test. This would have follow-on effects in terms of making the differences between the various estimates of the deceleration parameter smaller~\cite{Hubble-arXiv, Hubble-PRD}, which would give us greater confidence in the reliability and robustness of the conclusions.

The fact that there are such large differences between the cosmological parameters deduced from the different models based on physically plausible distance indicators should give one pause for concern. These differences do not arise from any statistical flaw in the analysis, nor do they in any sense represent any ``systematic'' error, rather they are an intrinsic side-effect of what it means to do a least-squares fit --- to a  finite-polynomial approximate Taylor series --- in a situation where it is physically unclear as to which if any particular measure of ``distance''  is physically preferable, and which particular notion of ``distance'' should be fed into the least-squares algorithm.  (This ``feature'' --- some may call it a ``limitation'' --- of the least-squares algorithm in the absence of a clear physically motivated dynamical model  is an often overlooked confounding factor in data analysis~\cite{basic1, basic2, basic3, basic4, transformation, nonlinear}.)

\section{Systematic uncertainties}

In addition to the purely statistical uncertainties discussed above, one needs to make an estimate of the systematic uncertainties, and following NIST guidelines~\cite{NIST}, combine the statistical and systematic  uncertainties in quadrature
\begin{equation}
\sigma_\mathrm{combined} = \sqrt{ \sigma_\mathrm{statistical}^2 + \sigma_\mathrm{total-systematic}^2 }.
\end{equation}
Estimating systematic uncertainties is notoriously difficult. A careful description of our own preferred way of estimating systematic uncertainties is fully discussed in~\cite{Hubble-arXiv, Hubble-PRD}, wherein we consider both modelling and historical uncertainties. It should be emphasized  that the (to our minds) overly optimistic estimates of systematic uncertainties commonly found in the literature do not greatly change our conclusions below.

\section{Expanded uncertainties}

After due allowance is made for estimating the systematic uncertainties, the  NIST guidelines~\cite{NIST} recommend defining an ``expanded uncertainty'' by
\begin{equation}
U_k = k \; \sigma_\mathrm{combined}.
\end{equation}
Here the factor $k$ is chosen for scientific (or legal) reasons to be such that one is ``certain'' that the true result lies within the stated range. The tradition within the social and medical sciences is to accept $k=2$ (that is, two-sigma, corresponding approximately to 95\% confidence intervals) as being sufficient to draw valid conclusions.  Particle physics has traditionally adopted $k=3$ as the minimum standard for claiming ``evidence for'' a given hypothesis. (This is the origin of the aphorism: ``If it's not three-sigma, it's not physics''.)  Over the last 20 years or so, particle physics has moved to the more conservative consensus that $k=5$ is the minimum standard for claiming ``discovery'' of ``new physics''.  Our best estimates for the combined and expanded uncertainties are presented in tables \ref{T:q-combined}--\ref{T:j-combined}.

\begin{table}[!htdp]
\caption{Deceleration parameter summary: 
Combined and expanded uncertainties.}
\begin{center}
Deceleration parameter summary: 
Combined and expanded uncertainties.\\[5pt]
\begin{tabular}{|c|c|c|c|c|}
\hline
dataset & redshift & 
$q_0\pm\sigma_\mathrm{combined} $ & $q_0 \pm U_3$  & $q_0 \pm U_5$ \\
\hline
{\sf legacy05} & $y$ & $-0.66\pm0.42$  &$-0.66\pm1.26$ &$-0.66\pm2.10$\\
{\sf legacy05} & $z$ & $-0.62\pm0.23$ & $-0.62\pm0.70$ &$-0.62\pm1.15$\\
{\sf gold06}     & $y$ & $-0.94\pm0.39$  &$-0.94\pm1.16$ &$-0.94\pm1.95$ \\
{\sf gold06}     & $z$ & $-0.58\pm0.23$ & $-0.58\pm0.68$ & $-0.58\pm1.15$\\
\hline
\end{tabular}
\end{center}
\label{T:q-combined}
\end{table}%

\begin{table}[!htdp]
\vskip -20 pt
\caption{Jerk parameter summary: 
Combined and expanded uncertainties.}
\begin{center}
Jerk parameter summary: 
Combined and expanded uncertainties.\\[5pt]
\end{center}
\hskip-15pt
\begin{tabular}{|c|c|c|c|c|}
\hline
dataset & redshift & 
$(j_0+\Omega_0)\pm\sigma_\mathrm{combined}$ & $(j_0+\Omega_0)\pm U_3$   & $(j_0+\Omega_0)\pm U_5$\\
\hline
{\sf legacy05} & $y$ & $+2.65\pm4.63$ & $+2.65\pm13.9$  & $+2.65\pm23.2$\\
{\sf legacy05} & $z$ & $+1.94\pm1.72$ & $+1.94\pm5.17$ & $+1.94\pm8.60$\\
{\sf gold06 }    & $y$ & $+6.47\pm4.75$ &  $+6.47\pm14.2$ &  $+6.47\pm23.8$\\
{\sf gold06 }    & $z$ & $+2.03\pm1.75$ &  $+2.03\pm5.26$ &  $+2.03\pm8.75$\\
\hline
\end{tabular}

\label{T:j-combined}
\end{table}%

\section{Conclusions}

\enlargethispage{15pt}

What can we conclude from this? While the ``preponderance of evidence'' is certainly that the universe is currently accelerating, $q_0<0$, this is not yet a ``gold plated'' result, \emph{at least not without bringing in other physical assumptions and observations}; such as a specific dynamical model  [\eg, $\Lambda$CDM] and/or invoking knowledge of $\Omega_m$ or the CMB data, all of which are subject to their own additional theoretical assumptions.  It is certainly more likely that the expansion of the universe is accelerating, than that the expansion of the universe is decelerating --- but this is not the same as having definite evidence in favour of acceleration. 

\begin{quote}
\emph{We wish to emphasize the point that, regardless of one's views on how to combine formal estimates of uncertainty, the very fact that different distance scales yield data-fits with such widely discrepant estimates for the cosmological parameters strongly suggests the need for extreme caution in interpreting the supernova data.}
\end{quote}

There are a number of other more sophisticated statistical methods that might be applied to the data to possibly improve the statistical situation. For instance, ridge regression, robust regression, and the use of orthogonal polynomials and ``loess curves'' could all be adopted and adapted to focus more carefully on the region near redshift zero~\cite{basic1, basic2, basic3,  basic4, transformation, nonlinear}. However one should always keep in mind the difference between \emph{accuracy} and \emph{precision}~\cite{Bevington}. More sophisticated statistical analyses may permit one to improve the precision of the analysis, but unless one can further constrain the systematic uncertainties such precise results will be no more accurate than the current  situation. 

However, we are certainly not claiming that all is grim on the cosmological front --- and do not wish our views to be misinterpreted in this regard --- there are clearly parts of cosmology where there is plenty of high-quality data, and more coming in, constraining and helping refine our models.  But regarding some specific cosmological questions the catch cry should still be ``Precision cosmology? Not just yet"~\cite{precision}. In closing, we strongly encourage readers to carefully contemplate figures \ref{F:ln_dF_legacy05}--\ref{F:ln_dQ_gold06} as an inoculation against over-interpretation of the supernova data.  

\begin{quote}
\emph{Ultimately, it is the fact that  figures \ref{F:ln_dF_legacy05}--\ref{F:ln_dQ_gold06}  do \emph{not} exhibit any overwhelmingly obvious trend that makes it so difficult to make a robust and reliable estimate of the sign of the deceleration parameter.}
\end{quote}

Finally we remind the reader that it is the putative acceleration of the expansion of the universe, no matter how derived, that then (via the Freidmann equations), is taken to imply the existence of ``dark energy''.  In the absence of truly compelling model-independent evidence for cosmic acceleration one has to be at least a little cautious regarding the existence of ``dark energy''.

\enlargethispage{15pt}


\end{document}